\documentclass[epjCONF]{svjour}
\usepackage{graphicx}
\usepackage[varg]{txfonts} 
\usepackage[latin1]{inputenc}
\session-title{40$^{th}$ LIAC}
\begin{document}
\title{Hot subluminous Stars: Highlights from the MUCHFUSS and Kepler missions}
\author{U. Heber\inst{1}\fnmsep\thanks{\email{ulrich.heber@sternwarte.uni-erlangen.de}} \and S. Geier\inst{1} \and B. Gaensicke\inst{2}}
\institute{Dr. Karl Remeis-Observatory \& ECAP, Astronomical Institute, Friedrich-Alexander University Erlangen-Nuremberg, Sternwartstr. 7, (6049 Bamberg, Germany \and Department of Physics, University of Warwick, Coventry CV4 7AL, UK}
\abstract{Research into hot subdwarf stars is progressing rapidly. We present recent important discoveries. First we review the knowledge about magnetic fields in hot subdwarfs and highlight the first detection of a highly-magnetic, helium-rich sdO star. We briefly summarize recent discoveries based on Kepler light curves and finally introduce the closest known sdB+WD binary discovered by the MUCHFUSS project and discuss its relevance as progenitor of a double-detonation type Ia supernova.      
} 
\maketitle
\section{Introduction}
\label{intro}
Subdwarf stars of spectral type O and B (sdO, sdB) are core helium-burning stars at the hot end of the horizontal branch (the extreme horizontal brach, EHB) or have evolved even beyond that stage. About half of the sdBs reside in close binaries; companions are white dwarfs or low-mass main-sequence
stars. Binary population-synthesis models explain naturally the actual sdB
binary fractions if white-dwarf mergers are considered as well. 
Research into hot subdwarf stars is a florishing field because the wide variety of phenomena observed in hot subdwarf stars can be used to tackle important issues in modern astrophysics, ranging from the fate of planets around evolved stars to the progenitors of type Ia supernovae and the origin of the UV excess in early-type galaxies. A review of the field has been given by Heber \cite{heber09}, while the crucial issue of hot subdwarf formation is reviewed at this meeting by Geier \cite{geierliege}. A more recent census can be found in the proceedings of the {\it Fifth Meeting on Hot Subdwarf Stars and Related Objects} \cite{kilkenny12}. Here we highlight some important recent discoveries. First we address the issue of magnetic fields of hot subdwarf stars, because major progress has been achieved in 2012. As for many other fields of astrophysics, the Kepler mission has a great impact also on the development of the research field of hot subdwarfs by providing light curves of pulsating sdB stars and close sdB binaries. We shall give examples in section 3. Finally, we shall introduce the MUCHFUSS project and present the most recent highlight discoveriies including CD$-$30$^\circ$11223, a candidate supernova Ia progenitor system. 

\section{A magnetic subdwarf O star}
\label{sec:1}

\begin{figure}
  \begin{center}
  \begin{minipage}[b]{7.cm}
    \includegraphics[width=0.95\textwidth]{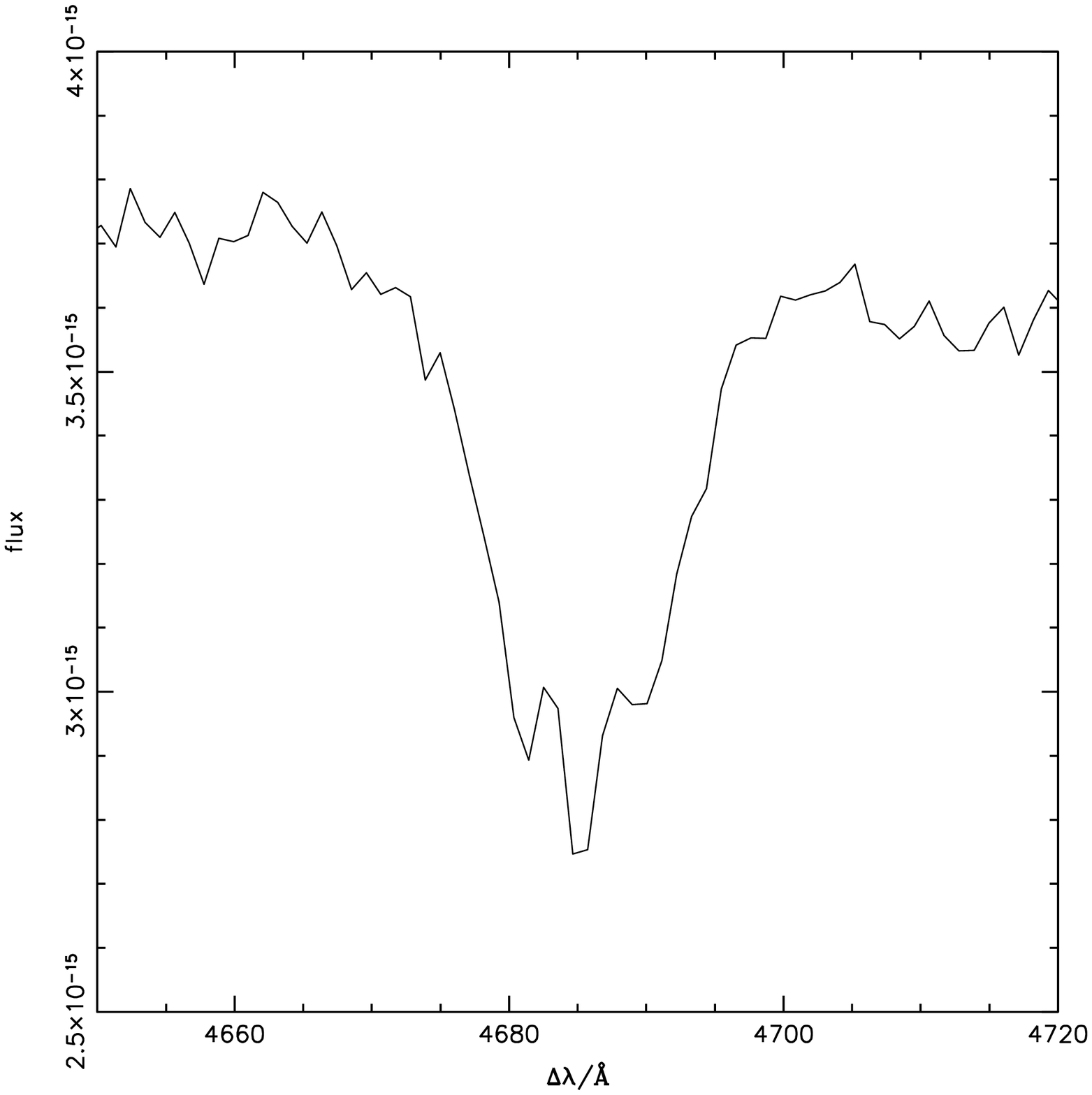}
  \end{minipage}
  \begin{minipage}[b]{7cm}
    \includegraphics[width=0.95\textwidth]{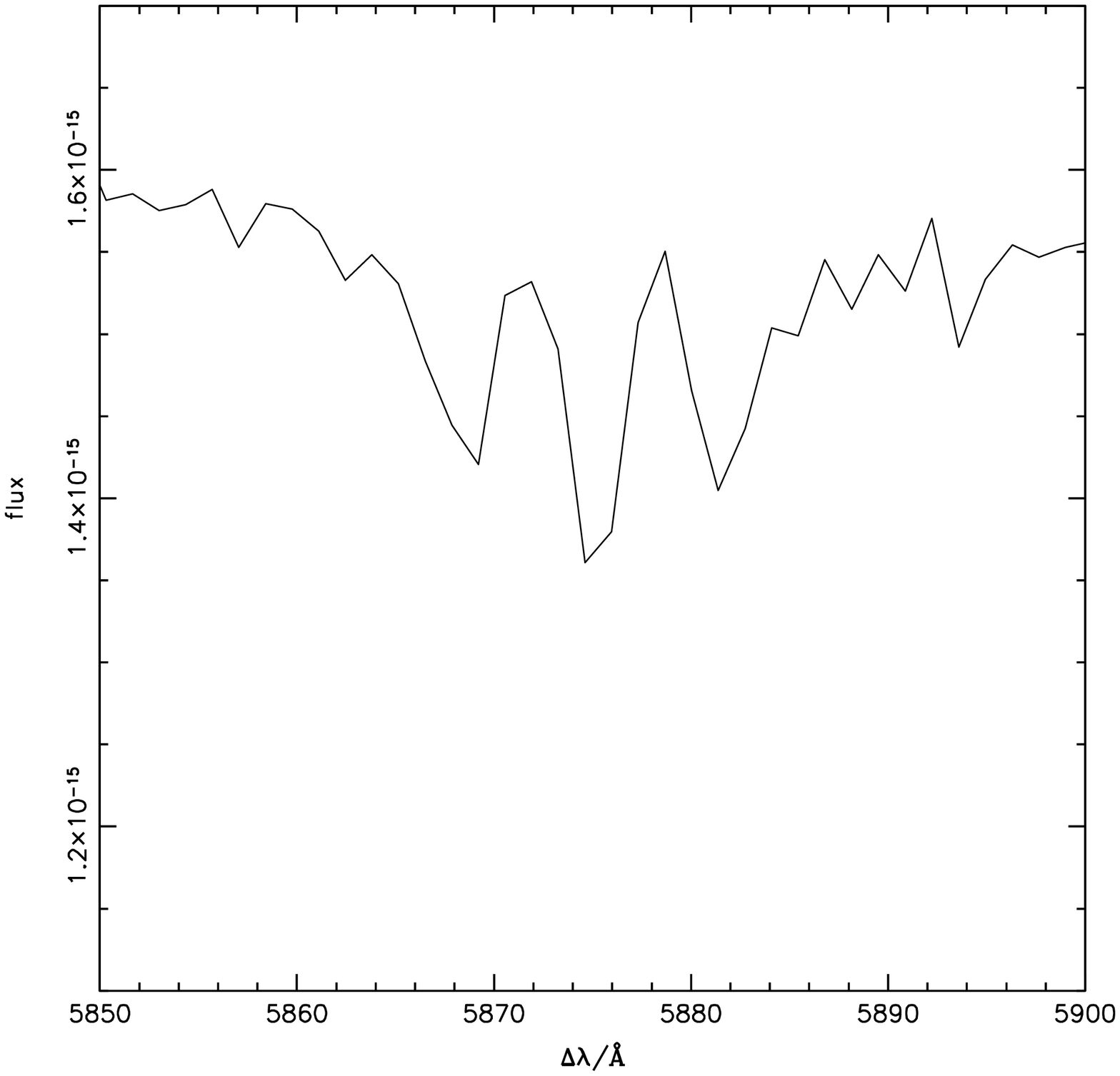}
  \end{minipage}
 \caption{Spectral lines of ionized (4686\AA, left hand panel) and neutral helium (5875\AA, right hand panel). The Zeeman splitting into three components is obvious, indicating a magnetic field strength of $\approx$300--700 kG}
 \label{mag}
\end{center}
\end{figure}

Attempts to detect  
magnetic fields in hot subdwarf stars up to now met with little success. Only a little more that half a dozen stars \cite{elkin96,elkin98,otoole05,chountonov12} 
 have been studied up to recently and the detection of kG-fields has been reported. More extensive studies have become available recently \cite{petit12,mathys12,landstreet12} covering 41 hot subdwarfs.
 Landstreet et al. (2012) carried out a critical reinvestigtion of published data from the FORS1 instrument at the ESO-VLT and new observations.  
In no case any field with standard errors of the
order of 200 -- 400 G could be found. Previous claims based on VLT-FORS data were found to be invalid because of wavelength calibration problems. Landstreet et al. (2012) conclude ''There is presently no strong evidence for the occurrence of a magnetic field in any sdB or sdO star, with typical longitudinal field uncertainties of the order of 2-400 G. It appears that globally simple fields of more than about 1 or 2 kG in strength occur in at most a few percent of hot subdwarfs.'' Therefore it came by some surprise when a helium-rich sdO star was recently discovered to show Zeeman splitting (\cite{gaensicke12}, see Fig.~\ref{mag}), indicating the presence of a $\approx$300 -- 700 kG magnetic field.

However, most magnetic white dwarfs are of spectral type DA, that is hydrogen rich. The newly discovered magnetic sdO star, however, is helium rich and therefore probably evolves into a white dwarf, that is also helium rich, that is into a DB white dwarf. However, highly magnetic DBs are rare compared to magnetic DAs. Hence, we should expect to find many more highly magnetic hot subdwarfs, which still need to be discovered.

\section{Hot subdwarf stars in the Kepler field}
\label{sec:2}
The Kepler mission provided light curves of unprecedented precision for hot subdwarf stars. The initial interest in the Kepler field sdB stars was to study pulsating sdB stars \cite{oestensen11}. Two classes of such multi-mode, low-amplitude pulsators are known, the 
short-period V361~Hya  stars (P$\approx$ 120 -- 600 s) and the V1093~Her stars (P$\approx$ 45 -- 120 min). The former are g-mode while the latter are p-mode pulsators. Both classes of pulsator are separated in the HRD, V361~Hya  stars are hotter than 28000\,K while all V1093~Her stars have temperatures below that limit. A few so-called hybrid pulsators are found near the dividing temperature. A longstanding puzzle is that the fraction of sdB stars in the (hotter) V361~Hya instability strip is very small while most of the sdBs in the (cooler) V1093~Her instability strip do pulsate. It was anticipated that the high precision and long duration of the Kepler light curves would allow to detect pulsations in many more sdB stars at photometric amplitudes lower than can be obtained from ground.

During the first year of the Kepler mission. Thirtytwo
sdB pulsator candidates hotter than 28 000\,K  have been surveyed \cite{oestensen11} and only one was found to be an unambiguous V361 Hya pulsator. Amongst the sixteen sdB stars of the Kepler sample cooler than 28000\,K, twelve stars (75\%) showed V1093~Her type pulsation, the fraction being in agreement with expectations from ground-based studies.
{\O}stensen et al. (2012) conclude that ''thanks to the exceptional precision of the Kepler measurements,
we can now conclude that there certainly are sdB stars, both on
the hot and on the cold ends of the EHB, that show no trace of
pulsations. Possible explanations for the non-pulsators would have
to answer why the pulsation driving mechanism is suppressed in
some EHB stars and not in other.''

Besides pulsating stars, some sdB+WD binaries were found in the Kepler sample.
In addition to elipsoidal variations caused by tidal distortions of the sdB star the Kepler light curves show variations caused by Doppler boosting. This allows to determine the radial velocity semi-amplitude from the light curve alone. 
A particularly impressive case is provided by the eclipsing binary KPD~1946+4340.  A photometric radial velocity amplitude of 168$\pm$4 $\rm km\,s^{-1}$ in excellent agreement with the spectroscopic one of K=164.0$\pm$1.9 $\rm km\,s^{-1}$ was derived \cite{bloemen11}. Even gravitational lensing has to be taken into account, because it is found to affect the depth of the eclipse at orbital phase 0.5. The analysis of radial velocity and Kepler light curves  allowed Bloemen et al. (2011) to derive the masses of both the sdB and the white dwarf. The mass of the sdB 0.47$\pm$ 0.03 M$_\odot$ is very close to the canonical EHB mass and predictions by population synthesis models \cite{han03}. The mass of the companion of (0.59$\pm$0.02 M$_\odot$) is typical for C/O white dwarfs.

\section{MUCHFUSS}
\label{sec:3} 
The project {\it Massive Unseen Companions to Hot Faint Underluminous Stars from SDSS (MUCHFUSS)} 
aims at finding hot subdwarf binaries with massive companions through optical spectroscopic and photometry \cite{geier11}. 

The SDSS spectroscopic database is the starting point of that survey. Hot subdwarf candidates were selected by applying a colour cut to SDSS photometry. All point source spectra within the colours $u-g<0.4$ and $g-r<0.1$ were selected and downloaded from the SDSS Data Archive Server\footnote{das.sdss.org}. By visual inspection $\simeq10\,000$ hot stars were selected and classified. The sample contains $1369$ hot subdwarfs. 
Subdwarf B stars with radial velocities lower than $\pm100\,{\rm km\,s^{-1}}$
were rejected to filter out such binaries with normal disc kinematics, by far the majority of the sample. Another selection criterion is the brightness of the stars. Most objects fainter than $g\approx 19\,{\rm mag}$ have been excluded.

However, it turns out that the MUCHFUSS selection strategy also allows to detect low-mass companions to sdBs in very close orbits. Two eclipsing sdB binaries with  brown dwarf companions have been found in the course of the MUCHFUSS photometric follow-up campaign \cite{geier11b,geier12b} and are discussed in the next subsection, while the shortest period sdB+WD system from MUCHFUSS is highlighted in section 4.2.   

\subsection{Subdwarf B plus brown dwarf systems}

Subdwarf B stars with low mass, non-degenerate companions are named after the prototype HW~Vir stars, if they are eclipsing. Such objects are rare. The MUCHFUSS project has discovered two such systems via photometry at the Mercator telescope and the CAHA-2.2m telescope equipped with BUSCA. The mass of the eclipsing sdB binary J082053.53+000843.4 of $0.045$ -- $0.068\,M_{\rm \odot}$ turned out to be lower than the hydrogen-burning limit ($0.07$ -- $0.08\,M_{\rm \odot}$ depending on metallicity). Hence this HW~Vir system hosts a brown dwarf companion\cite{geier11}. 

\begin{figure}
\begin{center}
  \includegraphics[width=8cm, angle=-90]{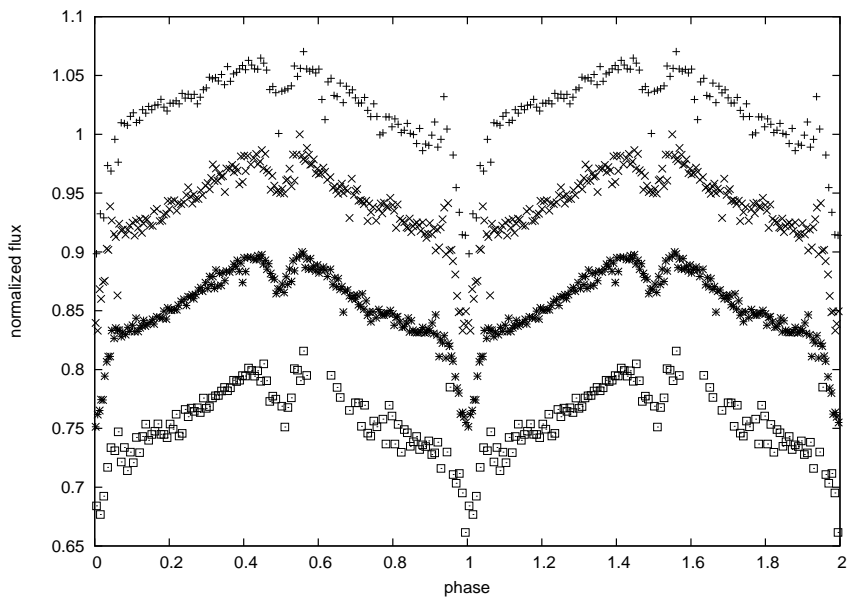}
\end{center}
\caption{Phased light curves of J162256.66+473051.1 taken with BUSCA (UV,B,R,IR-band). Primary and secondary eclipses can be clearly seen as well as the sinusoidal shape caused by the reflection effect (from \cite{geier12b}).}
\label{fig:J1622}
\end{figure}

Very recently a similar HW~Vir system, J162256.66+473051.1, was discovered in the course of the MUCHFUSS project (\cite{geier12b}, see Fig.~\ref{fig:J1622}). Its orbital period is as short as $\simeq0.07\,{\rm d}$ but the RV semi-amplitude is quite low ($\simeq47\,{\rm km\,s^{-1}}$). Although the analysis is still going on, it appears likely that the companion is substellar as well. 

MUCHFUSS's success to find low-mass companions illustrates that its target selection not only singles out sdB binaries with massive companions and therefore high RV-amplitudes, but also low mass systems with very short orbital periods. These results add to the growing evidence that low-mass stellar and substellar companions may play an important role in the formation of sdB stars (e.g. \cite{beuermann12}). 

\subsection{CD$-$30$^\circ$11223}
CD$-$30$^\circ$11223 is a sdB star recently discovered amongst blue stars in the GALEX survey \cite{vennes11}. The star was chosen as a bright backup target for our MUCHFUSS follow-up campaign. Due to bad observing conditions, which did not allow to observe the main targets, two medium resolution spectra ($R\simeq2200,\lambda=4450-5110\,{\rm \AA}$) were taken consecutively with the EFOSC2 spectrograph mounted at the ESO\,NTT at June 10, 2012. The spectra showed a very high radial velocity shift ($\simeq600\,{\rm km\,s^{-1}}$), which called for immediate follow-up to obtain time series spectroscopy and photometry.
A long
photometric time series from the SuperWASP planetary transit survey public archive was readily available and extensive time-series spectroscopy (e.g. with the William Herschel telescope on La Palma) were obtained on short notice.

\begin{figure}
\begin{center}
 \includegraphics[width=7.5cm]{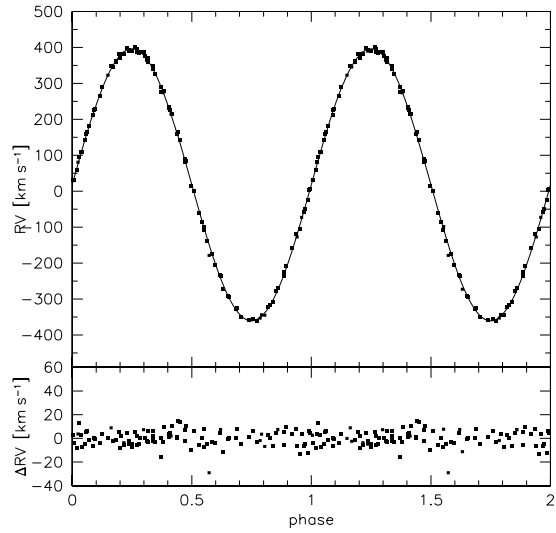}
 \caption{Radial velocity curve of CD$-$30$^\circ$11223 derived from 105 spectra taken with WHT/ISIS (from \cite{geier12}).}
\label{rv}
\end{center}
\end{figure}

\begin{figure}
\begin{center}
 \includegraphics[width=7.5cm,angle=90]{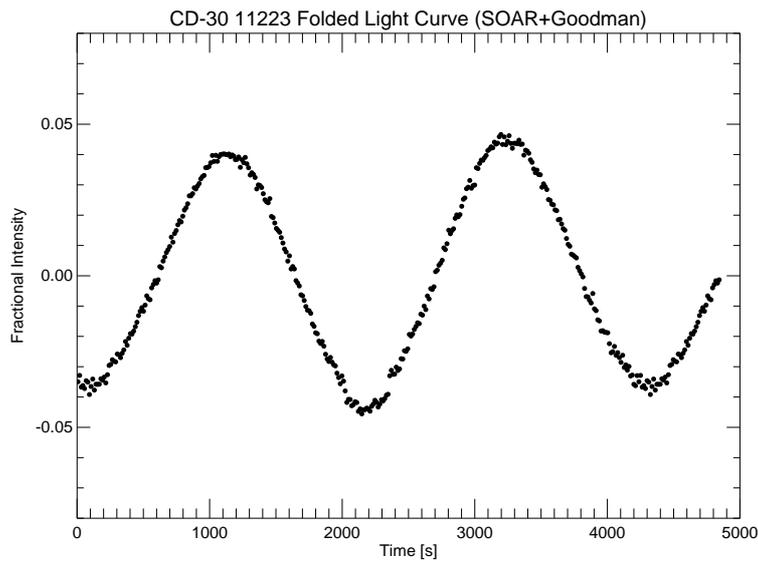}
 \caption{V-band light curve of CD$-$30$^\circ$11223 taken with SOAR/Goodman and folded to the orbital period (from \cite{geier12}).}
  \label{lc}
\end{center}
\end{figure}

While the visible primary is a typical sdB, the binary properties of this system are unique \cite{geier12}. The orbital period of only 1.17 hours, which is by far the shortest of any sdB binary and displays the largest radial velocity semi-amplitude of $376.6$ ${\rm kms^{-1}}$ (see Fig. \ref{rv}).
The orbital and atmospheric parameters of the sdB have been measured from time-series spectroscopy and allow to constrain the binary parameters (see Table 1). The SuperWASP light curve displayed variations due to the ellipsoidal deformation of the sdB star. This means that the companion must be a compact object, probably a white dwarf, the tidal drag of which causes the deformation of the sdB star. It was deemed likely that the system might be eclipsing. However, the SuperWASP data were of insufficient quality to judge. Therefore a new light curve was obtained with the SOAR telescope and eclipses were indeed detected (see Fig. \ref{lc}). An analysis of the optical light curve is ongoing and will constrain the system's inclination ($70-90^{\rm \circ}$). 
Combining all these results, the mass of the sdB ($>0.49\,M_{\rm \odot}$),
the mass of the WD companion ($>0.74\,M_{\rm \odot}$) will be constrained as well as the separation of the components ($0.6\,R_{\rm \odot}$); see Table 1 for other parameters of the system. Similar results have been derived by \cite{vennes12}.

The future evolution of this binary is particularly interesting since it can be a progenitor of an SN Ia via the
so-called sub-Chandrasekhar double-detonation scenario \cite{fink10}. In this scenario, the ignition of He-burning on the
surface of an accreting WD is predicted to trigger carbon-burning in the core even if the star is less massive than
the Chandrasekhar limit. 
The extremely short orbital period and small separation imply that the sdB star is close to fill its Roche-lobe
The surface gravity determined with the spectroscopic analysis of CD$-$30$^\circ$11223 implies a radius of  0.17$R_\odot$,  which corresponds
to 85\% of the Roche-lobe radius R$_{\rm Roche}\sim0.2R_\odot$.

Due to gravitational wave radiation the orbit will shrink and Roche-lobe overflow will start in about 30 Myrs, much shorter than the EHB life time of the sdB. Once about 0.1 M$_\odot$ of helium have accumulated on the C/O white dwarf a helium detonation will be launched that subsequently lits the C/O core.

\begin{table}
\label{compmasses}
\caption{\bf Parameters of the CD$-$30$^\circ$11223 system}
\begin{center}
\begin{tabular}{llll}
\hline
%
Distance & $d$ & [pc] & $290\pm50$\\
Visual magnitude$^{*}$ & $m_{\rm V}$ & [mag] & $11.90\pm0.18$\\
Effective temperature & $T_{\rm eff}$ & [K] & $29\,200\pm400$\\
Surface gravity & $\log{g}$           & & $5.66\pm0.05$\\
Helium abundance & $\log{y}$          & & $-1.50\pm0.07$\\
Projected rotational velocity & $v_{\rm rot}\sin{i}$ & [${\rm kms^{-1}}$] & $177\pm10$\\
Orbital period & $P$ & [d] & $0.0489790717\pm0.0000000038$\\
RV semi-amplitude & $K$ & [${\rm kms^{-1}}$] & $376.6\pm1.0$\\
System velocity & $\gamma$ & [${\rm kms^{-1}}$] & $19.5\pm2.0$\\
Binary mass function & $f(M)$ & [$M_{\rm \odot}$] & $0.27$\\
Subdwarf mass & $M_{\rm sdB}$ & [$M_{\rm \odot}$] & $>0.49$\\
Orbital inclination & $i$ & [$^{\rm \circ}$] & $67-90$ \\
Companion mass & $M_{\rm comp}$ & [$M_{\rm \odot}$] & $>0.74$\\
Separation & $a$ & [$R_{\rm \odot}$] & $0.6$\\
\hline\\
\end{tabular}
\end{center}
\end{table}

\section{Summary and conclusions}

We presented a mixed bag of selected highlights of pieces of research into hot subdwarf stars. The first discovery of a highly magnetic sdO star, the census of pulsating sdB stars in the Kepler field and the enigmatic sdB binaries discovered from Kepler light curves as well as by the MUCHFUSS project demonstrates that the field is florishing and progress is rapid.

\end{document}